\documentclass[12pt, a4paper,twoside,openright]{article}

\usepackage{latexsym}
\usepackage{indentfirst}
\usepackage{graphicx}
\usepackage{amsmath,amssymb,amsfonts,amsthm,amstext,amscd}
\usepackage{mathrsfs}
\usepackage{fancyhdr}
\usepackage{times}
\usepackage{enumerate}
\usepackage{natbib}
\usepackage[hang,flushmargin]{footmisc}

\fancypagestyle{plain}{%
\fancyhf{}%
\fancyhead[LO]{\textbf{Science \& Philosophy}
}
\fancyhead[RO]{Volume 10(2), 2022 }

\fancyfoot[CO,RE]{}
}

\begin{document}
\title{\textbf{An analysis of the concept of inertial frame in classical physics and special theory of relativity}}
\author{Boris Čulina\footnote{University of Applied Sciences Velika Gorica, Velika Gorica, Croatia; boris.culina@vvg.hr}}

\date{}

\maketitle

\begin{abstract}
\begin{normalsize}
\noindent
The concept of  inertial frame of reference in classical physics and special theory of relativity is analysed. It has been shown that this fundamental concept of physics is not clear enough.  A definition  of  inertial frame of reference is proposed which expresses its key inherent property. The definition is operational and powerful. Many other properties of  inertial frames follow from the definition, or it makes them plausible. In particular, the definition shows why physical laws obey space and time symmetries and the principle of relativity, it resolves the problem of clock synchronization and the role of light in it, as well as the problem of the geometry of inertial frames.\\
\textbf{Keywords}:  inertial frame of reference; space and time symmetries; the principle of relativity; clock synchronization; physical geometry \footnote{\noindent Received on August 19th, 2022. Accepted on December 9th, 2021. Published on December 31th, 2022. doi: 10.23756/sp.v8i2.827. ISSN 2282-7757; eISSN 2282-7765. \copyright The Authors. This paper is published under the CC-BY licence agreement.}
\end{normalsize}
\end{abstract}

%
%

\newpage

\section{Introduction}
The concept of inertial frame is a fundamental concept of physics. The opinion of the author is that not enough attention has been paid to such a significant concept, not only in textbooks, but also in the scientific literature. In the scientific and philosophical literature, many  issues related to the concept of inertial frame have been addressed, but, as far as the author is aware, a systematic analysis of this concept has not been made. DiSalle's article (\cite{sep-spacetime-iframes}) in the Stanford Encyclopedia of Philosophy gives an overview of the historical development of the concept of an inertial frame as an essential part of the historical development of physics. Thus, DiSalle's article is complementary to this article in its purpose and content. In this article, the concept of  inertial frame of reference is analysed only within the framework of classical physics and special theory of relativity. This analysis could contribute to the analysis that has yet to be done: the analysis of the concept of inertial frame of reference in general relativity and especially in quantum physics.

The first part of this article identifies the basic properties of inertial frames in classical physics and special theory of relativity. The second part of the article gives a definition of  inertial frame from which most other properties of inertial frames follow or this definition makes them plausible.

\section{Analysis}

\subsection{Newton}

In Philosophiæ Naturalis Principia Mathematica,  one of Newton's goal is to describe absolute motion. This description also includes relative motion:\footnote{English translation:  \cite{Principia}}

\begin{quote}
	\small
	
	Absolute, true and mathematical time, of itself, and from its own nature flows equably without regard to anything external, and by another name is called duration: relative, apparent and common time, is some sensible and external (whether accurate or unequable) measure of duration by the means of motion, which is commonly used instead of true time ...
	
	Absolute space, in its own nature, without regard to anything external, remains always similar and immovable. Relative space is some movable dimension or measure of the absolute spaces; which our senses determine by its position to bodies: and which is vulgarly taken for immovable space ...
	
	Absolute motion is the translation of a body from one absolute place into another: and relative motion, the translation from one relative place into another ...
	
\end{quote}

\noindent In this description Newton assumes that the geometry of absolute space is Euclidean geometry.

With his first law, the law of inertia, Newton describes the absolute motion of the body:

\begin{quote}
	\small
	
	Every body perseveres in its state of rest, or of uniform motion in a right line, unless it is compelled to change that state by forces impressed thereon.
\end{quote}

\noindent The same is true for the other two Newton laws. However, Newton shows that these laws also apply to reference frames that move uniformly with respect to the absolute frame.

\begin{quote}
	\small
	The motions of bodies included in a given space are the same among themselves, whether that space is at rest, or moves uniformly forwards in a right line without any circular motion.
\end{quote}

\noindent Today we call these frames inertial frames. Newton assumes that Euclidean geometry applies to them as well as to absolute space.

Newton's description of space and time provides a clear basis for his laws. These are absolute laws of absolute motion. But over time it has become clear that such an approach is untenable because it invokes ``phantoms'': absolute space and absolute time.\footnote{``In brief, Newton’s absolute space is a phantom that should never be made the basis of an exact science.'' (\cite{Lange})} However, inertial frames remain as frames in which these laws apply. But how to define them when absolute space and absolute time are gone? Furthermore,  from Newton we  inherit the hypothesis that the centre of mass of the world rests in the absolute frame (Book 3 Hypothesis I), so that the centre of mass of the solar system, since it is far from other masses, moves uniformly relative to the centre of the world.\footnote{ Immediately after Hypothesis I Newton makes a stronger claim, Proposition XI, that the centre of mass of the solar system also rests in absolute space. However, the assumptions stated in the proof are incomplete for such a conclusion.} Thus, we can connect an inertial frame with the centre of mass of the solar system. This system can be well experimentally approximated by the requirement that fixed stars have a constant position in it. When we refer to the solar system as a reference frame below, we will mean this frame. Now, inertial frames can be defined as frames that move uniformly or are at rest relative to this frame. Experiments show that, with some limitations, Newton's laws as well as Euclidean geometry are valid in such frames.

\subsection{Lange}

The first constructive critiques of Newton's conception of  inertial frame based on the concepts of absolute space and absolute time appear in the second half of the 19th century. Lange (\cite{Lange}) gives the following description of an inertial frame:\footnote{ English translation: \cite{Pfister}}

\begin{quotation}
	\small
	
	Definition I. An ``inertial system''\footnote{The terms ``inertial system'' and ``inertial timescale'' come from Lange.} is any coordinate system of the kind that in relation to it three points $ P $, $ P' $ , $ P''  $, projected from the same space point and then left to themselves – which, however, may not lie in one straight line – move on three arbitrary straight lines $ G $, $ G' $, $ G'' $ (e.g., on the coordinate axes) that meet at one point.
	
	Theorem I.\footnote{In Lange's text, the word \textit{theorem} has the meaning of a postulate.} In relation to an inertial system the path of an arbitrary fourth point, left to itself, is likewise rectilinear.
	
	Definition II. An ``inertial timescale'' is any timescale in relation to which one
	point, left to itself (e.g., P), moves uniformly with respect to an inertial system.
	
	Theorem II. In relation to an inertial timescale any other point, left to itself, moves
	uniformly in its inertial path.
	
\end{quotation}

Lange defines a coordinate inertial frame as a frame in which three free particles released from a single point move in non-collinear straight lines. His definition assigns an experimentally verifiable condition but is not constructive in the sense that it does not give how to construct such a frame. Lange then postulates that all free particles in such a frame move in  straight lines. Inertial time is defined as the time at which such a particle travels the same distance at the same time. This is the global time of an inertial frame and requires a measure in the geometry of the space of the inertial frame. Lange assumes Euclidean geometry. Again, the definition gives an experimentally verifiable condition but does not give a construction of such a time. Lange postulates that any other free particle travels the same distance in the same inertial time. The premise of the whole description is the existence of free particles and our ability to identify them. 

Lange gives a successful analysis of the assumptions of Newton's first law. However, the basis of his approach is to single out the frame of measuring space and time according to how things will look in it, as a frame in which the motion of a free particle is the simplest -- it is a uniform motion along a straight line. As Wheeler would say: ``Time is defined so that motion looks simple.''(\cite{Wheeler}). Although Lange assumes the concept of a straight line and Euclidean geometry, we could add to his analysis that space is defined so that a free particle moves along a straight line. 

In the same spirit is another analysis of the concept of  inertial frame given by Thomson (\cite{Thomson}).  He defines an inertial frame as a frame in which the bodies affected by the forces move according to Newton's laws and expresses the law of inertia as the assertion to the existence of such a frame. So, here too, an inertial frame is determined by how things will look in it.

\subsection{Modern textbooks}

Modern textbooks of classical mechanics (not including the special theory of relativity) generally define an inertial frame in one of the following ways that we can relate to Newton's and Lange-Thomson's approach.

\begin{enumerate}
	\item \textit{The empirical approach}. An inertial frame is a frame that moves uniformly with respect to the solar system. It is postulated that Newton's laws apply in this frame (it is sufficient to postulate this for one such frame).
	\item \textit{The convenient approach}. An inertial frame is a frame in which Newton laws apply. Most often only the first law of inertia is mentioned, and the others are postulated. It is also postulated that the solar system is such a frame, as well as frames that move uniformly relative to it (it is enough to postulate it only for the solar system).
\end{enumerate} 

Although the concept of inertial frame is a fundamental concept, as a rule it is not analysed in modern textbooks of classical mechanics -- the textbooks  start from the concept in the development of mechanics. The internal structure of an inertial frame is not analysed, especially the mechanism of measuring space and time in such a frame. It is simply assumed, more often implicitly than explicitly, that space is Euclidean, and time is global. The empirical approach does not analyse why Newton laws would be valid in an inertial frame but states it as an experimentally confirmed statement. In the convenient approach, Newton laws are valid by definition. However, this definition is practically useless because, for example,  we should  examine the motions of all free particles with all velocities in all directions to determine whether Newton's first law is valid. This definition of  inertial frame makes the term empirically unverifiable and does not show us how to construct such a frame. Therefore, as far as inertial frames are concerned, modern textbooks are a step backwards compared to Newton and Lange. Newton, using the concepts of absolute space and time, explains why his laws apply in inertial frames (because these frames move uniformly relative to absolute space) and why the solar system is inertial (it moves uniformly with respect to the centre of the world which is the absolute frame). Lange, in addition to bringing to light the important concept of inertial time, gives an empirically verifiable definition of  inertial frame including  inertial time in it.

In addition to the assumptions about an inertial frame that its space is Euclidean, time is global and absolute, and that Newton's laws apply, modern  textbooks of classical mechanics sometimes assume, more often implicitly than explicitly, that in an inertial frame space is homogeneous and isotropic and time homogeneous and directed. There is no explanation as to why this would be the case (often it is not explained clearly enough what that means). Among the exceptions, the well-known Landau-Lifshitz textbook (\cite{Landau}) should be singled out -- they define  inertial frame using symmetries. In search of a frame ``in which the laws of mechanics take their simplest form'' they opt for a frame ``in which space is homogeneous and isotropic and time is homogeneous''. Such a frame they call inertial.  Apart from the claim that such a frame ``can always be chosen'' and that ``there is not one but an infinity of inertial frames moving, relative to one another, uniformly in a straight line'', it is not stated how to operationally find such a frame. Furthermore, they assume Euclidean geometry and global time in such a frame. From this definition they derive Newton's first law, Lagrangian of a free particle, restrictions on the form of Lagrangian  of a closed system, and conservation laws, thus showing that such a definition of  inertial frame is very powerful.

The Landau-Lifshitz approach, which emphasizes the symmetries of space and time in an inertial frame, also belongs to convenient approaches that characterize an inertial frame as the frame in which the laws of mechanics are the simplest. Unlike this type of definitions that determine an inertial frame by how mechanical processes look in such a frame, a definition can be found in textbooks according to which an inertial frame is defined  by its inherent property: it is a frame such that  there are no external forces acting on it. However, neither such a description is sufficiently precise nor are the corresponding consequences drawn from the definition. For a typical example, we can cite a passage from Wikipedia (\cite{wiki}):

\begin{quote}
	\small
	
	In classical physics and special relativity, an inertial frame of reference is a frame of reference that is not undergoing acceleration. In an inertial frame of reference, a physical object with zero net force acting on it moves with a constant velocity (which might be zero) -- or, equivalently, it is a frame of reference in which Newton's first law of motion holds. An inertial frame of reference can be defined in analytical terms as a frame of reference that describes time and space homogeneously, isotropically, and in a time-independent manner. Conceptually, the physics of a system in an inertial frame have no causes external to the system.
\end{quote}

\noindent If we understand the first statement as a definition of an inertial frame, we have a typical situation in this approach: various properties of an inertial frame are listed, and they are in no way related to the definition.

\subsection{The special theory of relativity}

The special theory of relativity has brought key improvements in the conception of inertial frame. In his groundbreaking work (\cite{Einstein}) Einstein starts from the established concept  of  inertial frame: it is a frame in which ``the equations of mechanics hold good''. Thus, he accepts all classical assumptions, first of all Euclidean geometry which he considers realized by means of an extended solid body and rigid rods. However, in this paper, Einstein introduces two essential innovations related to inertial frames. The first is the generalization of the principle of relativity: not only are the laws of classical mechanics the same in all inertial frames, but all the laws of physics are the same in all inertial frames. Another innovation is the analysis of the concept of time in an inertial frame. Einstein starts from the fact that time is measured locally -- with the same clock. He assumes that in an inertial frame at each point in space we can have identical clocks that we need to synchronize to get the global time of the inertial frame. Einstein describes the synchronization of clocks at different places $ A $ and $ B $ in an inertial frame as follows:\footnote{English translation: \cite{Lorentz}}

\begin{quotation}
	\small
	
	\noindent We have so far defined only an ``$ A $ time'' and a ``$ B $ time''. We have not defined a common ``time'' for $ A $ and $ B $, for the latter cannot be defined at  all unless we establish by definition\footnote{This part of the translation is wrong and should read: ''\ldots and the latter can now be determined by establishing by definition\ldots''(J.D.Norton).} that the ``time'' required by light to travel from $ A $ to $ B $ equals the ``time'' it requires to travel from $ B $ to $ A $. Let a ray of light start at the ``$ A $ time'' $t_A$ from $ A $ towards $ B $, let it at the ``$ B $ time'' $t_B$ be reflected at $ B $ in the direction of $ A $, and arrive again at $ A $ at the ``$ A $ time'' $t'_A$.
	
	In accordance with definition the two clocks synchronize if
	
	\begin{displaymath}t_B-t_A=t'_A-t_B. \end{displaymath}
	
	We assume that this definition of synchronism is free from contradictions, and possible for any number of points; and that the following relations are universally valid:—
	
	\begin{enumerate}
		\item  If the clock at $ B $ synchronizes with the clock at $ A $, the clock at $ A $ synchronizes with the clock at $ B $.
		\item If the clock at $ A $ synchronizes with the clock at $ B $ and also with the clock at $ C $, the clocks at $ B $ and $ C $ also synchronize with each other.
	\end{enumerate}

	Thus with the help of certain imaginary physical experiments we have settled what is to be understood by synchronous stationary clocks located at different places, and have evidently obtained a definition of ``simultaneous'', or ``synchronous'',  and of  ``time''. The ``time'' of an event is that which is given simultaneously with the event by a stationary clock located at the place of the event, this clock being synchronous, and indeed synchronous for all time determinations, with a specified stationary clock.
	
	In agreement with experience we further assume the quantity
	
	\begin{displaymath}\frac{2{\rm AB}}{t'_A-t_A}=c, \end{displaymath}
	
	\noindent to be a universal constant — the velocity of light in empty space.
	
	It is essential to have time defined by means of stationary clocks in the stationary system, and the time now defined being appropriate to the stationary system we call it ``the time of the stationary system''.
	
\end{quotation}

In short, Einstein a) gave the definition of synchronization of two clocks by light, b) postulated that all clocks of an inertial frame can be consistently synchronized in the sense that synchronization of two clocks is an equivalence relation with exactly one equivalence class, c) that once synchronized clocks remain synchronized, and d) that the two-way speed of light (the speed measured on the same stationary clock) is the universal constant of  an inertial frame. 

Having thus obtained the global time of an inertial frame (``stationary system'', in Einstein's words), Einstein can define the concept of one-way velocity in an inertial frame and state his second postulate (the light principle):

\begin{quote}
	\small
	
	Any ray of light moves in the ``stationary'' system of co-ordinates with the determined velocity $ c $, whether the ray be emitted by a stationary or by a moving body. Hence
	\begin{displaymath}{\rm velocity}=\frac{{\rm light\ path}}{{\rm time\ interval}} \end{displaymath}
	where time interval is to be taken in the sense of the definition in § 1.\footnote{Einstein refers here to ``the time of the stationary system'' previously described.}
	
\end{quote}

Although Einstein, with his generalized principle of relativity and the light principle based on the analysis of the concept of time in an inertial frame, revolutionized physics, some things remained insufficiently clarified in the key part of his article quoted above:

\begin{enumerate}
	\item The problem of the conventionality of the definition of synchronization (\cite{Reichenbach}). Every definition of synchronization which is of the form $t_{\rm B} = t_{\rm A} + \varepsilon (t'_{\rm A}-t_{\rm A})$, where $0 < \epsilon <1$,  is in accordance with the principle of causality.   Is Einstein’s choice $\varepsilon = \dfrac{1}{2}$ physically different from other choices or is it just a pleasant convention with no physical significance? This problem has generated controversy that is still present today (\cite{Anderson, Jammer, sep-spacetime-convensimul}).
	
	\item The problem of consistent synchronization (in Einstein's sense) of all clocks. What properties of light are needed to achieve this? In particular, what is the role of the postulate of the constancy of the two-way speed of light in clock synchronization?
	
	\item The problem of possible circularity. Light signals are used for synchronization in order to express the light principle about light with the help of such synchronized clocks. For example, from the very definition of synchronization it follows that the one-way speed of light in opposite directions is the same. If we add to this the postulate of the constancy of the two-way speed of light, we get the light principle as a consequence of synchronization and not as an additional postulate. Thus, although Einstein introduces the clock synchronization procedure to articulate the light principle, in his work it remains unclear which properties of light are required for synchronization. The problem of circularity also occurs at a deeper conceptual level because Einstein uses clock synchronization to define the global time of an inertial frame. However, he describes an inertial frame as a frame in which ``the equations of mechanics hold good''. These laws contain the law of inertia, which presupposes the global time of an inertial frame in its formulation, and which Einstein's clock synchronization has yet to establish.

\end{enumerate}


In \cite{Minkowski}, Hermann Minkowski  gave the formulation of the special theory of relativity in terms of a certain structure in the space of events. In short, the Minkowski event space is a 4-dimensional affine space in which worldlines of free particles and light are special types of straight lines (timelike and lightlike straight lines), and in which the metric tensor is given  that is directly related to light signalling and time measurement by means of free-moving clocks. It is an elegant mathematical reformulation of the special theory of relativity that does not introduce essentially new elements into the concept of an inertial frame. We can understand Minkowski space as the structure in the event space generated by the structure of an inertial frame in a way that is invariant to the choice of an inertial frame. The light principle and the principle of relativity  are automatically built into this structure (the principle of relativity as  a condition on the physical laws that they must be formulated in terms of  Minkowski space). Conversely, inertial frames can be understood as decompositions of Minkowski space to which the structure of Minkowski space is isomorphically transferred. In such a decomposition the space of an inertial frame is still Euclidean and the decomposition itself corresponds to Einstein's clock synchronization (\cite{Malament}). However, in Minkowski's formulation the inherent property of an inertial frame becomes more visible. Namely, the worldlines of free particles are timelike straight lines in that space, so each inertial frame  is identified as a class of all mutually parallel timelike straight lines. If we imagine that each such straight line is a worldline of a free particle, and not a particle acted upon by forces in equilibrium, then an inertial frame is a class of all free material particles that are at rest with each other. Thus, an inertial frame in this space of events naturally appears as a frame by which we can identify all events  and whose main feature is that it is free, that its elements do not enter any interactions.


\subsection{The general theory of relativity and quantum physics}

Here we will dwell only on some general observations on the possibility of extending the above analysis to general relativity and quantum physics.

As is well known, the general theory of relativity sets physical limits on the classical concept of an inertial frame. Regardless of how we describe an inertial frame, the essential concept is the concept of  free particle, the concept that is incompatible with the ubiquity of gravity and must be reformulated into the notion of a free-falling particle. Thus, the classical concept of inertial frame can only be realized approximately, within a limited space and time. Nevertheless, it is the key idealization of the general theory of relativity, the ``infinitesimal'' element of which the whole theory is composed. Note that even in the general theory of relativity, inertial frames have a natural inherent description: they are free-falling frames.

Although inertial frames are an essential element of quantum description of the world, they are rarely explicitly mentioned in quantum physics textbooks. If they are mentioned, they are not analysed, but their properties from classical (non-quantum) physics are simply transferred. In Bohr's approach (\cite{Howard,Tanona}),  they are a macroscopic element that is an integral part of the quantum description of the world and is usually related to macroscopic measuring instruments. In such an approach, the classical concept of an inertial system retains its importance. However, in other approaches the concept loses its meaning. For example, in the approach described in  (\cite{Aharonov,Angelo_2011}) an inertial frame itself must be a quantum mechanical system. Then some classical properties of an inertial frame must be reformulated. For example, the property that it is a frame in which free particles move uniformly in straight lines is transformed into the property that the expected value of the position of a free particle changes uniformly along a straight line -- the property that is difficult to verify experimentally. On the other hand, the characterization that an inertial frame  is a frame on which nothing acts still makes sense. Furthermore, in quantum physics, space and time symmetries can be attributed to an inertial frame, as well as the principle of relativity. However, the quantum mechanical properties of an inertial frame make the basic purpose of such a frame problematic -- to identify when and where something happened. What kind of such identification does quantum physics enable, that is, what kind of structure does it bring into the space of events? In particular, what geometry does it introduce into the space of an inertial frame? These are the key unanswered questions (\cite{Penrose}):

\begin{quote}
	\small

	I do not believe that a real understanding of the nature of elementary
	particles can ever be achieved without a simultaneous deeper understanding of the nature of space-time itself. 
	
\end{quote}

\subsection{Properties of an inertial frame}

After this review, let us summarize which properties are attributed to inertial frames:

\begin{enumerate}
	\item A frame in which the centre of mass of the solar system is at rest and in which the fixed stars have a constant position is an inertial frame.
	\item The space and time of an inertial frame are such that  free particles in it move uniformly in straight lines. In general, it is a frame in which Newton's laws apply.
	\item An inertial frame is a frame in which the laws of physics have the simplest form.
	\item An inertial frame is a frame on which there are no external forces. Or even more restrictively, it is a frame composed of free particles (there are no external or internal interactions) that are at rest with each other.
	\item The space of an inertial frame is Euclidean.
	\item The space of an inertial frame is homogeneous and isotropic, and time is homogeneous and directed.
	\item The time of an inertial frame is local, the local times can be synchronized and so the global time of an inertial frame can be obtained.
	\item Frames which move uniformly in  straight lines relative to an inertial frame are inertial frames and there are no other inertial frames.
	\item The principle of relativity: The laws of physics are the same in all inertial frames.
	\item The light principle: The speed of light in vacuum is the same in all inertial frames, regardless of the mode of light formation.
\end{enumerate}

\section{Definition and consequences}
\subsection{Definition of the concept of inertial frame} 

For a successful definition of a property, it is not enough that, in addition to formal correctness, the definition is only extensionally correct -- objects that have the defined property are precisely those objects that   we want to single out  from some multitude of objects. Such is, for example, Plato's definition of man as a two-legged animal without feathers. The most important criterion that the definition should meet is to be intensionally correct -- to single out objects according to some of their essential properties. Unlike the first two criteria, we are not yet able to give this third most important criterion a sufficiently precise form.\footnote{The criterion of extensional correctness has as precise a form as it is clear to us on an extensional level which objects we want to single out.}  But this does not mean that in particular cases we cannot distinguish better from worse definitions. Of course, in the choice of a definition, the criteria of precision (how precise the terms we use to define a new term) and effectiveness (how effectively we can examine whether an object has a defined property) are important, too. Of the properties listed in the previous section, inertial frames are characterized in an extensional sense by properties 1) together with 8), and properties 2), 3) 4) and 6). 

Criterion 1) plus 8) is an experimental determination. Thus, its meaning is poor, and we cannot relate it to other properties of an inertial frame. We can only postulate them independently.

Criteria 2), 3) and 6) identify an inertial frame by how physical processes look in it. These are external characterizations of an inertial frame that cannot explain its other properties. In addition, these criteria are not operational -- they do not show how to find such a frame. Criterion 3), in addition to being imprecise (what does it mean to have the simplest form?), provides no basis for identifying such frames. Since we do not know all the laws of physics, we cannot  know in which frame they have the simplest form. Moreover, it is possible that some laws have the simplest form in one type of frame and other laws  in another type of frame.  Criterion 2) is clear because it is limited to Newton's laws. But accepting this criterion would mean an unnecessary limitation of the concept of an inertial frame to classical Newtonian physics. The necessary universality can be obtained only if criterion 2) is limited to the description of the motion of a free particle. The main purpose of the reference frame is to identify where and when something happened, and the requirement  that in an inertial frame a free particle  moves rectilinearly and uniformly, is precisely the requirement for the space and time of the frame. But such a requirement is too weak to be related to other properties of an inertial frame. If we want to reinforce it with other requirements for space and time, primarily space and time symmetries, then  we come to criterion 6). However, an inertial frame must have inherent physical characteristics that affect what the laws of physics look like in it and not to be adjusted so that in it those laws have a certain form. Furthermore, this second approach only makes sense if we can formulate physical laws independently of the concept of a reference frame, which is operationally questionable. This is true for space and time symmetries, too. Such an external characterization  only makes sense if we can describe this space and time structure independently of the concept of reference frame. For example, in \cite{Brehme} inertial frames are defined as frames that are isomorphic to the Minkowski space.  However, the structure of Minkowski space is operationally derived from the structure of inertial frames, so this definition is only an elegant mathematical solution until we give Minkowski space a direct physical interpretation. This interpretation must explain why physical laws must be formulated in Minkowski space, that is, why they must have space and time symmetries, as well as satisfy the principle of relativity. However, even if we were to achieve such a definition of an inertial frame, structurally we would obtain a characterization of an inertial frame that it is a frame that (due to an isomorphism) has the structure of a Minkowski space. But again, it is an external characterization that does not tell us why an inertial frame would have such a space and time structure. Also, the definition would not be operational.

Criterion 4) is the only inherent criterion, a criterion that mentions the properties of the reference frame itself. While the aforementioned characterizations identify an inertial frame by how we describe physics in it, this characterization determines an inertial frame by what happens to the frame itself. Thus, in terms of  intensional correctness, it is the best criterion. It is also an operational criterion, unlike criteria 2) 3) and 6). In addition, it is very powerful. When we clarify the basic idea that an inertial frame is a free frame, a frame on which nothing acts, we will get a definition of  inertial frame from which almost all the remaining listed properties of an inertial frame can be derived or at least made plausible. For properties that cannot be related to the concept of  inertial frame, it will be shown that there are good reasons why, by their nature, they do not fall under the concept. Therefore, we will take criterion 4) to define inertial frame.

We will call \textit{reference frame}  any frame that allows us to  identify events spatially and temporally. The same reference frame can provide multiple coordinate systems for the identification. For example, the Euclidean space is a reference frame for determining position, and various coordinate systems for identification can be defined in it. Thus, we will distinguish a reference frame from the  coordinate reference frame that can be built in it. An inertial frame will be a special type of a reference frame.

The condition that there are no external forces on an inertial frame is too weak. If we look at a solid body that is not affected by external forces, it can rotate. This rotation is registered by the appearance of internal tensions in the body. However, the condition that we do not allow internal forces in an inertial frame is too strong. Since an inertial frame must provide spatial and temporal determination of events, it must contain certain measuring instruments. Therefore, we will allow the existence of localized closed parts within which there is an interaction, but not the existence of a non-localized interaction, such as interactions caused by rotation, that could disrupt the symmetries of an inertial frame. This does not preclude the existence of large-scale solid bodies in an inertial frame, because in the absence of external forces and rotation we can ignore internal tensions in the body. There can only be isolated and localized interactions.

Since an inertial frame serves to determine the space and time coordinates of an event, it must also have the ability to determine that its parts are at rest relative to each other. Only localized deviations from rest in closed processes that serve to measure space and time are allowed.

Based on the above considerations, we define  \textit{inertial frame} as a reference frame such that the following holds:

\begin{enumerate}
	
	\item There are no external forces on the frame.
	\item Interactions within the frame are possible only in localized and closed parts of the frame.
	\item Parts of the frame are at rest, except for possible localized deviations from rest. 
\end{enumerate}

The precision of this definition is limited by the precision of the terms used in it, but we will show that it is precise enough to be usable.

This definition does not follow from Newton's description of inertial frames as frames that move uniformly relative to the absolute frame. The law of inertia states that free frames move uniformly relative to the absolute frame, but the reverse is not true: frames that move uniformly relative to the absolute frame do not have to be free -- these include frames that are affected by forces in equilibrium.

Likewise, this definition does not follow from the standard definition of an inertial frame as a frame in which free particles move uniformly in a straight line or are at rest. Parts of such a frame are at rest in the frame but this does not mean that they are free -- this includes parts that are affected by external or internal non-localized forces that are in equilibrium.

This definition of  inertial frame is one of the standard definitions of  inertial frame, somewhat more precise here than usual. It is suggested by Newton's approach and by the standard definition through the observation of a free particle, but it is more restrictive than these definitions, as shown in the previous paragraphs. Such a definition occurs naturally from the aspect of event space (whether it has a Galilean structure or a Minkowski structure), as well as from the aspect of general relativity, where it corresponds to free-falling reference frames, if we localize them in space and time  enough.

The most important term on which the definition of an inertial frame rests is the concept of interaction. Thus, for the definition to be operational, it assumes that we know what kind of interactions exist. However, other definitions are based on this same concept, too. For example, the same assumption lies behind the concept of a free particle in Newton’s first law. Ultimately, we can understand this definition as a working definition, which changes every time we discover new interactions. 

If we assume that the interactions decrease with distance then we can consider that any frame that is far enough away from other bodies and in which there are no non-localized interactions is approximately inertial. Thus, the solar system (a system in which the centre of mass is at rest, and which has a constant direction with respect to fixed stars) can be considered inertial. Since external actions as well as rotation cause non-localized tensions, we can experimentally check whether a reference frame is inertial with an appropriate system of accelerometers and gyroscopes. For Einstein, an inertial frame is tied to the extended rigid body. If there are no external actions on the rigid body and it does not rotate, then its parts are free and in  a constant mutual position, so it determines an inertial frame. A free observer with a clock and theodolite, which sends light signals around and measures the time of sending and receiving signals   also forms an inertial frame. Of course, the definition of  inertial frame formulated here is an idealization in relation to which we can estimate how much the actual frame of reference corresponds to an inertial one. As already mentioned, the most significant restriction on the realization of such frames is set by the general theory of relativity, but also by quantum physics.

\subsection{Space and time symmetries of an inertial frame}

Since an inertial frame is free, the space of this frame is homogeneous and isotropic. Any inhomogeneity and non-isotropy would mean the existence of  external forces or an unnecessary internal symmetry breaking (e.g., to choose a different unit of measure in each direction). Likewise, any time inhomogeneity would mean the presence of external forces or unnecessary internal symmetry breaking, so such a space is also time homogeneous. These inherent symmetries of an inertial frame can be extended to measurements of space and time  but also to the description of all closed processes in such a frame. By definition, a \textit{closed system} has no interaction with the environment, so the events in it are independent of the space and time in which it is located. When such a system is observed from an inertial frame in which all points, directions and time moments are equal, then in an inertial frame such a system can be described in such a way that it has the specified symmetries. Thus, we can set \textit{the symmetry principle} of describing physical processes in an inertial frame: \textit{In an inertial frame the laws of physics for closed systems have space homogeneity and isotropy as well as time homogeneity}. This principle derives not only from how physical processes take place but also from how we can describe them in an inertial frame. Note, we do not have to describe a closed physical process that way. But an inertial frame gives us the ability to describe them that way, and it is to be expected that such a description is the best in every respect. I would also note that it makes sense to state this principle even before we have an elaborate structure of space and time. Moreover, we can and must (if we want to exploit the advantages of an inertial frame) apply it to the very determination of the geometry of space and the structure of time in an inertial frame. The approach to the description of physical processes in which we try to preserve the original symmetries of an inertial frame I will call \textit{the inertial frame approach}\label{ifa}.

\subsection{Space and time of an inertial frame}

The simplest closed system is a system composed of one (massive) particle -- it is a free particle. The ray of light, if we assume the absence of ether, is also a simple closed system.

Each closed periodic process, i.e., the process that returns to the initial state, including the initial position, determines the local measurement of time at that place. This is the general definition of \textit{local clock}. It could be an atomic clock. It can also be a free particle or light that bounces off something and returns to its starting position (Langevin clock). For a unit of time we can take some standardized process, for example a free particle created in the standard way that bounces off something or a light particle created in the standard way that bounces off something, if we assume that it is a closed system (that there is no ether). Here we do not have to assume that no matter how we create light it always gives the same unit of time (the light principle). Due to space and time symmetries, all closed periodic processes must measure the same time up to the choice of the unit of measure and their operation is independent of position, orientation, and elapsed time. Let us show more precisely that this is so. Let us have two closed periodic processes (two clocks) $ C_1 $ and $ C_2 $ to measure time. In general, the relation between the times $ t_1 $ and $ t_2 $ of the duration of a process measured  by the clocks $ C_1 $ and $ C_2 $ is a continuous function $ f $: $ t_2 = f (t_1) $. We will show that this function is a direct proportion: $t_2 = a\cdot t_1$, for some $a>0$. For this purpose, we will consider the time $ a $ of the duration of the periodic process $ C_1 $ measured on the clock $ C_2 $. Measured on the clock $ C_1 $ it is equal to 1. Thus $ a = f (1) $. Measuring by clock $ C_2 $ the duration of $ n $ consecutive $ C_1 $ processes, due to time homogeneity, will give $na= f(n)$.This means that if measuring any process with the clock $ C_1 $ yields fraction $ \dfrac{n}{m} $ then the measurement with the clock $ C_2 $ will yield $ \dfrac{n}{m}a = f(\dfrac{n}{m}) $. Because of the continuity of the function $ f $, this means that $ x \cdot a = f(x) $, for any positive real number $ x $. Thus, the function $ f $ is a direct proportion. By adding spatial symmetries to this,  we have shown that \textit{all clocks at all points of an inertial frame are equal}.

Since free particles or rays of light, assuming the absence of ether, are the simplest examples of closed systems, their trajectories must be the simplest examples of curves in the space of an inertial frame. Due to the homogeneity and isotropy of space, such a trajectory must have the same spatial characteristics at each location. So, it is natural to take these paths to define  \textit{straight lines} in the space of an inertial frame. Distances can always be measured by the same standardized periodic process by which we locally define time -- by means of a standardized free particle or a standardized light (assuming no ether). If since the sending of the standardized free particle (or light) from the point $ A $, its rejection from the point $ B $ and its return to the point $ A $ the elapsed time $ t $ is measured on the clock in $ A $,  then we can take that time for the measure of distance. But due to the isotropy of space, it is more natural to take half of that time to measure \textit{distance}:  $d(A,B) = \dfrac{t}{2}$. This does not change anything significantly because the measurement is always determined up to the multiplicative factor. Note that in this way we can also check an important element of the definition of  inertial frame, that the parts of the frame are at rest. After the reflection of a particle or light, there is a displacement of the body from which the reflection is made. However, in an inertial frame, such localized deviations from rest are allowed by definition, provided that this shift is subsequently reversed. \textit{Due to space and time symmetries, any choice of standardized free particles or standardized light (assuming no ether) gives the same geometry up to the  unit of measure}. Locally, we can make these measurements more conveniently using rigid rods. Due to space and time symmetries, such a measuring instrument, as well as the measuring system generated by it, can be reproduced at any point and in all directions. And, due to the symmetries, this leads to the same geometry. The geometry of the space of an inertial frame, the geometry in which the paths of free particles and light rays are straight lines, and in which the distance measurement is based on the described measurement of elapsed local time, is homogeneous and isotropic -- all points are equal and all directions are equal. Although scale symmetry is not generally valid for closed processes, we will show that it is valid for the geometry of an inertial frame. We will show that Thales's basic proportionality theorem holds: for the lengths of the segments marked in the figure below it holds that if $ x' = \alpha x $ and $ y '= \alpha y $ then $ z' = \alpha z $.

\begin{figure}
	\begin{center}
	\includegraphics[scale = 0.15]{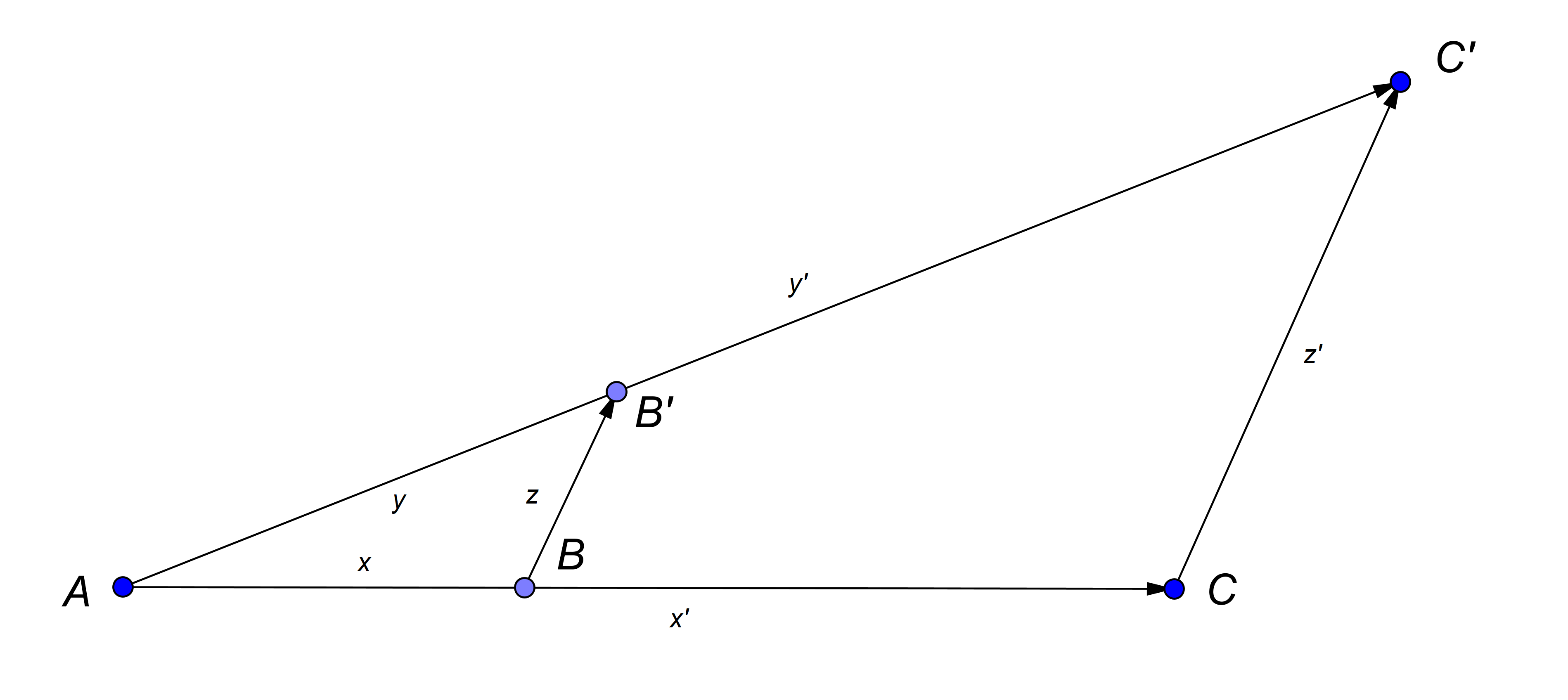}

	$ x=\overline{AB},\  y=\overline{AB'},\  z=\overline{BB'},\  x'=\overline{AC},\  y'=\overline{AC'},\  z'=\overline{CC'}$
	\end{center}
		\caption{Thales's proportionality theorem}
\end{figure}

\noindent This is due to the equality of all clocks in an inertial system. Let $ x $, $ y $ and $ z $ lengths be measured using a periodic closed process $ C_1 $ to which a unit measure is equal to $ c_1 $: $x = t_xc_1$, $y = t_yc_1$ and $z = t_zc_1$. Imagine another periodic closed process $ C_2 $ such that its unit of measure is equal to $c_2 = \alpha c_1$. Measuring $ x'$ and $ y' $ with the clock $ C_2 $ gives the same numerical value as measuring $ x $ and $ y $ with the clock $ C_1 $:  $x' = t_xc_2$ and $y' = t_yc_2$. Since the geometry is determined by measuring time, and all clocks are equal, just as the  numbers measured with the clock $ C1 $ determine the triangle $ x-y-z $, so the same numbers measured with the clock $ C_2 $ determine the triangle $ x'-y'-z' $. That's why it has to be $z' = t_zc_2$. Therefore, $z'=\alpha z$. This proves Thales's theorem. In \cite{Cu3} it is shown that from the assumptions of homogeneity, isotropy and scale symmetry of space 
axioms of Euclidean geometry can be obtained. This means that the geometry of an inertial frame is  Euclidean geometry.

The symmetries of space and time solve both the problem of clock synchronization (including the question of conventionality) as well as the problem of possible circularity of the description. Using free particles or light, we can synchronize clocks in an inertial frame with the same procedure we used to determine the measurement of distances -- we send a standardized free particle or light (assuming no ether) from one clock to the next and back. Symmetries give us the freedom to choose the means of synchronization. We can use any standardized free particle (we standardize the way of generating its motion) or a standardized ray of light, assuming that the light is a closed system (that there is no ether). If we assume that the motion of light is independent of the source of origin (the light principle), then we do not have to standardize light at all. \textit{Due to the above symmetries, whatever standardized process we use, we will always get the same clock synchronization} (if we synchronize them in one way, we will find for every other way, that it gives the same synchronization). If we synchronized the clock $ B $ with the clock $ A $ and that the synchronized clocks will remain synchronized, we will denote $ A \textrm { sinc } B $. 

Even before synchronization,  time homogeneity of an inertial frame tells us something about the connection of the time read by the clocks at the places $ A $ and $ B $. If we sent the standard signal from the clock at $ A $ in the moments $ t_1 $ and $ t'_1 $, and the clock at $ B $ received them in the moments $ t_2 $ and $ t'_2 $, then the difference in elapsed time is on both clocks same: $ t'_1-t_1 = t'_2-t_2 $. This is equivalent to the condition that the difference in signal travel time read on the $ B $ clock on arrival and on the $ A $ clock on  departure is always the same: $ t_2-t_1 = t'_2-t'_1 $. We will call this property of clocks  \textit{time homogeneity of clocks} in an inertial frame. 

\begin{figure}
	\begin{center}
	\includegraphics[scale = 0.7]{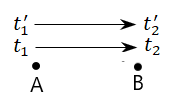}
	
	$t'_1-t_1 = t'_2-t_2 $
	
	$ t_2-t_1 = t'_2-t'_1 $
	\end{center}
	\caption{time homogeneity of clocks}
\end{figure}

If we look at all possible synchronizations that are in accordance with the principle of causality, they are of the form $ t_ {2} = t_ {1} + \varepsilon (t_1, A,B)(t_3-t_1) $, $ 0 <\varepsilon (t_1,A,B) <1 $, where $ t_1 $ is the time read on the clock at $ A $ when sending the signal from  $ A $, $ t_2 $ is the time read on the clock at $ B $ when the signal arrives at  $ B $, and $ t_3 $ is the time read on the clock at $ A $ when the signal returns to  $ A $. Due to the time homogeneity, $ \varepsilon (t_1, A,B) $ must not depend on $ t_1 $ and due to space symmetries it must not depend on $ A $ and on the direction towards $B$ -- it must be the same number $\varepsilon$ for all points. In particular, it must be the same number to synchronize the clock at $ A $ with the clock at $ B $. We can get this synchronization by reflecting the previously described signal once again back to the point $ B $ where its arrival will be read at the moment $ t_4 $ on the clock at $ B $.

\begin{figure}
\begin{center}
	\includegraphics[scale = 0.7]{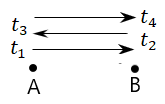}
	
\end{center}
\caption{Synchronizing clocks}
\end{figure}
\noindent Synchronizing the clock at $ B $ with the clock at $ A $ gives

$$t_{2} = t_{1} + \varepsilon (t_3-t_1)$$

\noindent To get the relationship $ t_3 $ with $ t_2 $ and $ t_4 $, we will eliminate $ t_1 $ in the above relationship using  time homogeneity of the clocks:

$$t_3-t_1 = t_4 - t_2$$

\noindent We will get 

$$t_{3} = t_{2} + (1-\varepsilon) (t_4-t_2)$$

\noindent Due to space isotropy, it must be $ 1- \varepsilon = \varepsilon $, or $ \varepsilon = \dfrac{1}{2} $. The conclusion is that the symmetries of an inertial frame require that the synchronization relation is symmetric, that is, that $\varepsilon = \dfrac{1}{2}$. Other choices would break the symmetry. Thus, the concept of  inertial frame  leads to Einstein’s clock synchronization and not some other synchronization. This choice is not just a matter of convention, but it is part of the inertial frame approach (page \pageref{ifa}) to the study of nature. The situation is the same as when setting symmetry conditions on physical laws. So here too, an inertial frame allows us to choose Einstein's synchronization, and we must certainly take advantage of this in the study of nature -- to  keep the symmetries of inertial frame, and so to choose Einstein synchronization. In what follows, we will mean by synchronization precisely this symmetrical Einstein synchronization.

It is not difficult to show that time  homogeneity of clocks is equivalent to the condition that once synchronized clocks remain synchronized, and the symmetry of a synchronization relation is equivalent to the condition that $\varepsilon = \dfrac{1}{2}$.


With Einstein's synchronization, we can synchronize all clocks with one clock in a symmetrical way. However, due to  space and time symmetries of an inertial frame, we will get the same result no matter what clock we take for the synchronizing clock. Thus, an inertial frame realizes Einstein's assumption of consistency of synchronization. We can show this in more detail. It follows from the isotropy of space that if we send a signal from point $ A $ so that it comes to point $ B $, it bounces to point $ C $, from where it bounces back to point $ A $, the time $t_{\leftarrow}$  to return to point $ A $ (measured at the clock at $ A $) will be equal to the time $t_{\rightarrow}$ it takes for the signal to go around these points in the opposite direction: from $ A $ through $ C $ and $ B $ back to $ A $ (measured at the clock at $ A $). This condition was considered by Reichenbach and called the roundabout axiom (\cite{Reichenbach}).

\begin{figure}

\begin{center}
	\includegraphics[scale = 0.7]{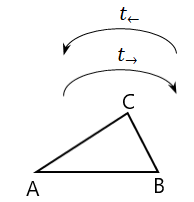}
	
	$ t_{\leftarrow} = t_{\rightarrow}$
\end{center}
\caption{the roundabout axiom}
\end{figure}
\noindent It is easy to show (\cite{Reichenbach,Mac}) that, assuming time homogeneity of clocks, the circular isotropy of the synchronization signal is equivalent to the transitive property of Einstein synchronization. Thus, synchronization is also a transitive relation. So, we got that synchronization is an equivalence relation (reflexivity is trivial -- each clock is synchronized with itself). Since we can synchronize all other clocks with one clock, this means that this equivalence relation has only one class, that is, that \textit{we have a consistent synchronization of all clocks in the sense that every two clocks of an inertial frame are synchronized.}
%
%
%
%
%
%
%
%
%

We can now say that after the described synchronization procedure, all  clocks of an inertial frame show the same time -- the \textit{global time of the inertial frame}. This time is an inertial time because it follows from the invariance of synchronization to the choice of a standardized particle or light for the synchronisation procedure that the time satisfies the Lange condition: a free particle travels the same distance at the same time.

Now that we have measures of space and time in an inertial frame, we can measure in it the (one-way) velocities of all free particles as well as the light produced in all possible ways.

Note that in this system of choice of units of space and time, a standardized free particle or standardized light has a velocity equal to 1 (both two-way and one-way velocity) -- during  time  $ t $ it travels  the distance $ t $. Of course, we can have another  system of measurement of distances. If the system respects the symmetries of an inertial frame, we will get the same geometry. Only the unit of measurement of distance will be different. Such is, for example, the standard system of measurement with rigid rods.

We see that \textit{Einstein synchronization of an inertial frame can be obtained without the use of light and so independently, without any circularity, the light postulate can be set}. Even if we choose one standardized light for the synchronization procedure (assuming that there is no ether, i.e., that the light is a closed system), this does not mean that every light, regardless of the conditions of its origin, has the same two-way speed as the standard light. The light principle, that every light has the same two-way speed $C$, is an additional postulate that goes beyond the concept of an inertial frame -- it does not follow from it.  The one-way light principle is then a consequence of the two-way light principle and clock synchronization. Of course, that speed is the same in all inertial frames. If we also use light to measure distance, this speed is equal to 1.

We see that a proper understanding of an inertial frame solves all the synchronization problems that arise in Einstein’s article. A nice logical analysis of the synchronization problem and the role of light in it without assumptions of space and time symmetries can be found in \cite{Mac,Min}.


Let us point out at the end that this concept of  inertial frame says nothing about the direction of time. The existence of the direction of time is ubiquitous and inertial frames only inherits this property. Thus, this property is independent of the concept of inertial frame.

\subsection{The principle of relativity}

Since free particles move in an inertial frame uniformly in straight lines, and an inertial frame is a frame composed of free particles with a constant relative position, inertial frames also move uniformly in a straight line (in the sense that all its parts move uniformly in parallel straight lines while maintaining their mutual position) relative to a given inertial frame. Thus, all inertial frames are in relative uniform motion with each other. As already stated, the reversal is not valid. If a frame of reference moves uniformly in a straight line relative to an inertial frame, this does not mean that it is inertial, i.e.,  that its parts move freely. Such motion can also occur in the presence of external forces that are mutually in equilibrium. Their presence disturbs space and time symmetries of the frame and all previous analysis and the one that follows loses its basis. Here, this conception of inertial frame differs from the standard one according to which any frame that moves uniformly relative to an inertial frame is also an inertial frame.

The considerations we have applied to establish  space and time symmetries of a closed physical process in an inertial frame, can also be applied to establish the possibility of the description in an inertial frame of a closed process that is invariant to the choice of an inertial frame. By definition, a closed system has no interaction with the environment, so the events in it are independent of an inertial frame from which we observe it. Since inertial frames do not differ  in the way of observing a closed system then such a system can be described in a way invariant to the choice of an inertial frame. Thus, we can establish \textit{the principle of relativity} of the description of physical processes in an inertial frame: \textit{The laws of physics for a closed system are invariant to the transition from one inertial frame to another}. This principle, as well as the principles of space and time symmetries, derives not only from how physical processes take place but also from how we can describe them in an inertial frame. I repeat, we do not have to describe a closed physical process like that. But inertial frames give us the ability to describe them that way, and it is to be expected that such a description is the best in every respect. I think that with these considerations I have supplemented Geroch in \cite{EinsteinP}, page 179, who says, among other things:

\begin{quote}
	\small
	
	The principle of relativity, then, hides within itself a
	subtle distinction—between what is and what is not
	taken as a law of physics. Indeed, it could be argued that
	a better perspective is to regard the principle of relativity, not as a general principle of nature at all, but rather
	as a guideline for distinguishing between those phenomena that are to be taken as “laws of physics” and
	those that are not. Phenomena that have the same
	description in every frame -- that is, phenomena that are
	compatible with the principle of relativity -- are to be
	accorded the status of physical laws, while phenomena
	that have different descriptions in different frames are to
	be regarded as merely specific phenomena. This is not a
	purely philosophical distinction: It can have consequences as to how physics is conducted. 
\end{quote}

It is not sufficiently known that the principle of relativity follows from space and time symmetries of inertial frames (see, for example, the proof in \cite {Rindler}, page 40). It follows directly from this result that  the principle of relativity is founded in the same way as space and time symmetries.

\section{Conclusion}

Although the definition of the concept of inertial frame formulated here may seem insufficiently precise and ``fragile'', we see that it leads to a very robust and powerful properties of inertial frames. It ensures the existence of space and time symmetries of an inertial frame, as well as the principle of relativity. These symmetries together with the principle of relativity place certain restrictions on the possible laws of physics that guide us in finding them. Also, powerful conservation laws follow from them. Thanks to the symmetries of an inertial frame, all clocks are equivalent to each other (robustness!) -- they define the same time up to a unit of measure. Symmetries also ensure time homogeneity of the clocks at various locations in an inertial frame, which is equivalent to the condition that  once synchronized clocks remain synchronized. Also, symmetries ensure that we use free particles or light (assuming that once the light is emitted it is a closed system -- no ether) to synchronize the clocks, and again in a robust way -- no matter which procedure we choose to synchronize we will always get the same synchronization -- Einstein clock synchronization. Likewise, no matter which procedure we choose to measure space, we always get the same space geometry -- Euclidean geometry.

From the concept of inertial frame defined herein, we derived or at least made plausible all the enumerated properties of inertial frames with the following exceptions, which have been shown to remain clearly separated from the concept of inertial frame:

\begin{enumerate}
	\item Although all inertial frames move uniformly with each other, it is not necessary that a frame of reference that moves uniformly relative to an inertial frame is also an inertial frame.
	\item The speed of light in vacuum is the same in all inertial frames, regardless of how it is formed (the light principle).
	\item Time has a direction.
\end{enumerate}

\bibliographystyle{abbrvnat}
\bibliography{Inertial_frames_1}

\end{document}